\documentclass{ifacconf}
\usepackage{amsmath,mathptmx,amsfonts}
\usepackage{scalefnt}
\usepackage{graphicx}
  \graphicspath{{./pics/}}
   \DeclareGraphicsExtensions{.pdf}
\usepackage{xspace}
\usepackage{natbib}
\usepackage{tikz}
  \usetikzlibrary{arrows,automata,shapes,backgrounds}
  \usetikzlibrary{decorations.pathmorphing}
  \usetikzlibrary{calc}

\newcommand{\eps}{\varepsilon}
\DeclareMathOperator{\A}{\mathcal A}

\DeclareMathOperator{\D}{\mathcal D}

\newcommand{\complclass}[1]{{\sc #1}\xspace}
\newcommand{\NL}{\complclass{NL}}

\newcommand{\coNP}{\complclass{coNP}}
\newcommand{\NP}{\complclass{NP}}
\newcommand{\PSpace}{\complclass{PSpace}}
\newcommand{\PTime}{\complclass{PTime}}

\usepackage[textwidth=1.9cm,color=green!10,textsize=footnotesize]{todonotes}

\begin{document}
\begin{frontmatter}

\title{On Opacity Verification for Discrete-Event Systems}

\author[upol]{Ji\v{r}\'{i} Balun}
\author[upol,mu]{Tom{\' a}{\v s}~Masopust}

\address[upol]{Faculty of Science, Palacky University in Olomouc, Olomouc, Czechia}
\address[mu]{Institute of Mathematics of the Czech Academy of Sciences, Brno, Czechia\\ (e-mail: jiri.balun01@upol.cz, masopust@math.cas.cz)}

\begin{abstract} 
  Opacity is an information flow property characterizing whether a system reveals its secret to an intruder. Verification of opacity for discrete-event systems modeled by automata is in general a hard problem. We discuss the question whether there are structural restrictions on the system models for which the opacity verification is tractable. We consider two kinds of automata models: (i) acyclic automata, and (ii) automata where all cycles are only in the form of self-loops. In some sense, these models are the simplest models of (deadlock-free) systems. Although the expressivity of such systems is weaker than the expressivity of linear temporal logic, we show that the opacity verification for these systems is still hard.
\end{abstract} 

\end{frontmatter}

\section{Introduction}
  In practical applications, it is desirable to keep some information about a system secret. Such a requirement results in additional restrictions on the information flow. Several information flow properties studied in the literature include {\em anonymity\/} of~\cite{SchneiderS96}, {\em noninterference\/} of~\cite{Hadj-Alouane05}, {\em secrecy\/} of~\cite{Alur2006}, {\em security\/} of~\cite{Focardi94}, and {\em opacity\/} of~\cite{Mazare04}.

  Opacity is the property whether a system prevents an intruder from revealing the secret. The intruder is modeled as a passive observer with the complete knowledge of the structure of the system but with only limited observation of its behavior. Based on its observation, the intruder estimates the behavior of the system, and the system is opaque if the intruder's estimation never reveals the secret. In other words, for any secret behavior of the system, there is a non-secret behavior that looks the same to the intruder. 
  
  If the secret is modeled as a set of secret states, the opacity is referred to as {\em state-based opacity}. \cite{Bryans2005} introduced state-based opacity for systems modeled by Petri nets, \cite{SabooriHadjicostis2007} adapted it to systems modeled by (stochastic) automata, and \cite{BryansKMR08} generalized it to transition systems. 
  If the secret is modeled as a set of secret behaviors, the opacity is referred to as {\em language-based opacity}. Language-based opacity was introduced by \cite{Badouel2007} and \cite{Bubreil2008}.
  Many researchers studied opacity from different perspectives, including its verification and the synthesis of opacity-ensuring controllers. For more details, we refer the reader to the overview by~\cite{JacobLF16}.

  Several notions of opacity have been discussed in the literature, e.g., current-state opacity, initial-state opacity, initial-and-final-state opacity, language-based opacity, $K$-step opacity, or infinite-step opacity. 
  Current-state opacity asks whether the intruder cannot, based on its state estimation, in any instance of time decide whether the system is currently in a secret state.
  Initial-state opacity asks whether the intruder can never reveal whether the computation started in a secret state. 
  Initial-and-final-state opacity of \cite{WuLafortune2013} is a generalization of both current-state opacity and initial-state opacity. The difference is that the secret is encoded as a pair of an initial and a marked state. Consequently, initial-state opacity is a special case of initial-and-final-state opacity where the marked states do not play a role. Similarly, current-state opacity is a special case where the initial states do not play a role.
  
  While current-state opacity is a property to prevent the intruder from revealing whether the current state of the system is a secret state, initial-state opacity prevents the intruder from revealing whether the system started in a secret state, that is, at any time during the computation.
  In more detail, the difference between initial-state opacity and current-state opacity is that initial-state opacity requires that the intruder cannot reveal the secret neither at the beginning of the computation nor in any state later during the computation, while current-state opacity only requires that the intruder cannot reveal the secret in the current state. It may, however, happen that the intruder may reveal that the system was in a secret state in the future. For instance, assume that the intruder estimates that the system is in one of two states, and that the system proceeds in the next step by an observable event that is possible only from one of the states. Then, the intruder reveals the state in which the system was one step ago. 
  
  This problem has been considered in the literature and led to the introduction of two notions: $K$-step opacity of \cite{SabooriHadjicostis2007} and infinite-step opacity of \cite{SabooriH09}. While $K$-step opacity requires that the intruder cannot reveal the secret in the current and $K$ subsequent states, infinite-step opacity requires that the intruder can never reveal that the system was in a secret state.

  The complexity of opacity verification has widely been investigated in the literature and is often based on the computation of observer. Thus the problem belongs to \PSpace. It is actually \PSpace-complete for most of the discussed notions. Indeed, \cite{Cassez2012} showed that the verification of current-state opacity is at least as hard as deciding universality, which is \PSpace-complete for nondeterministic automata as well as for deterministic automata with partial observation. 
  
  However, \PSpace-completeness of universality requires a nontrivial structure of the model and the ability to express all possible strings. This give rise to a question whether there are structurally simpler systems for which the verification of opacity is tractable. We investigate the problem for, in our opinion, structurally the simplest systems: for acyclic automata (that do not have the ability to express all strings, and actually express only a finite number of strings) and for automata where all cycles are in the form of self-loops (which may still seem trivial in the structure, because as soon as the system leaves a state, it can never return to that state).

  In this paper, we study the effect of those structural restrictions on the verification of current-state opacity. Notice first that using the polynomial reductions of \cite{WuLafortune2013} among current-state opacity, initial-state opacity, initial-and-final-state opacity, and language-based opacity, allows us to deduce immediate consequences of our study for other notions of opacity as well. We discuss these consequences in detail in Section~\ref{secConsequences}. 
  
  Our contribution is as follows. We show that deciding language-based weak opacity for systems where both the secret and the non-secret languages are modeled by NFAs is \NL-complete (Theorem~\ref{thm_sdnl-c}). Then we show that deciding current-state opacity for deterministic finite-state systems with only three events, one of which is unobservable, is \PSpace-complete (Theorem~\ref{thm2}). Considering systems with only one observable event, we show that the complexity decreases to \coNP-complete (Theorem~\ref{thm3}). Then we study acyclic systems that have only a finite amount of different behaviors and show that deciding current-state opacity for acyclic systems with at least two observable events is \coNP-complete (Theorem~\ref{thm7}), and that the complexity decreases to \NL-complete in the case the systems have a single observable event (Theorem~\ref{thm7b}). Finally, we investigate the simplest deadlock-free systems, that is, systems where all cycles are self-loops, and show that whereas deciding current-state opacity of such systems with a single observable event is \NL-complete (Theorem~\ref{thm90}), the problem for systems having three events, one of which is unobservable, is \PSpace-complete (Theorem~\ref{thm10}) even for deterministic systems.

\section{Preliminaries}
  We assume that the reader is familiar with the basics of automata theory, see~\cite{Lbook} for details.

  For a set $S$, $|S|$ denotes the cardinality of $S$, and $2^{S}$ the power set of $S$. An alphabet $\Sigma$ is a finite nonempty set of events. A string over $\Sigma$ is a sequence of events from $\Sigma$. Let $\Sigma^*$ denote the set of all finite strings over $\Sigma$; the empty string is denoted by $\varepsilon$. A language $L$ over $\Sigma$ is a subset of $\Sigma^*$. The set of all prefixes of strings of $L$ is the set $\overline{L}=\{u \mid uv \in L\}$. For a string $u \in \Sigma^*$, $|u|$ denotes its length, and $\overline{u}$ the set of all prefixes of $u$.

  A {\em nondeterministic finite automaton\/} (NFA) over an alphabet $\Sigma$ is a structure $\A = (Q,\Sigma,\delta,I,F)$, where $Q$ is a finite  set of states, $I\subseteq Q$ is a set of initial states, $F \subseteq Q$ is a set of marked states, and $\delta \colon Q\times\Sigma \to 2^Q$ is a transition function that can be extended to the domain $2^Q\times\Sigma^*$ by induction. Equivalently, the transition function is a relation $\delta \subseteq Q\times \Sigma \times Q$, where, e.g., $\delta(q,a)=\{s,t\}$ denotes two transitions $(q,a,s)$ and $(q,a,t)$. For a state $q\in Q$, the language marked by $\A$ from $q$ is the set $L_m(\A,q) = \{w\in \Sigma^* \mid \delta(q,w)\cap F \neq\emptyset\}$, and the language generated by $\A$ from $q$ is the set $L(\A,q) = \{w\in \Sigma^* \mid \delta(q,w)\neq\emptyset\}$. The language marked by $\A$ is then the union $\bigcup_{q_0\in I} L_m(\A,q_0)$; similarly for the language generated by $\A$.

  The NFA $\A$ is {\em deterministic\/} (DFA) if $|I|=1$ and $|\delta(q,a)|\le 1$ for every $q\in Q$ and $a \in \Sigma$. For DFAs, we identify singletons with their elements and simply write $p$ instead of $\{p\}$. Specifically, we write $\delta(q,a)=p$ instead of $\delta(q,a)=\{p\}$.
  
  A {\em discrete-event system\/} (DES) $G$ over $\Sigma$ is an automaton (NFA or DFA) together with the partition of the alphabet $\Sigma$ into two disjoint subsets $\Sigma_o$ and $\Sigma_{uo}=\Sigma\setminus\Sigma_o$ of {\em observable\/} and {\em unobservable events}, respectively. In the case where all states of the automaton are marked, we simply write $G=(Q,\Sigma,\delta,I)$ without specifying the set of marked states. 
  
  The opacity property of a DES   is based on partial observation of events described by the projection $P\colon \Sigma^* \to \Sigma_o^*$, which is a morphism defined by $P(a) = \varepsilon$ for $a\in \Sigma_{uo}$, and $P(a)= a$ for $a\in \Sigma_o$. The action of $P$ on a string $\sigma_1\sigma_2\cdots\sigma_n$ with $\sigma_i \in \Sigma$ for $1\le i\le n$ is to erase all events that do not belong to $\Sigma_o$; namely, $P(\sigma_1\sigma_2\cdots\sigma_n)=P(\sigma_1) P(\sigma_2) \cdots P(\sigma_n)$. The definition can readily be extended to languages.

  A {\em decision problem\/} is a yes-no question. A decision problem is {\em decidable\/} if there is an algorithm that solves it. Complexity theory classifies decidable problems to classes based on the time or space an algorithm requires to solve the problem. The complexity classes we consider are \NL, \PTime, \NP, and \PSpace denoting the classes of problems solvable by a nondeterministic logarithmic-space, deterministic polynomial-time, nondeterministic polynomial-time, and deterministic polynomial-space algorithm, respectively. The hierarchy of the classes is \NL $\subseteq$ \PTime $\subseteq$ \NP $\subseteq$ \PSpace. Which of the inclusions are strict is an open problem. The widely accepted conjecture is that all are strict. A decision problem is \NL-complete (resp. \NP-complete, \PSpace-complete) if (i) it belongs to \NL (resp. \NP, \PSpace) and (ii) every problem from \NL (resp. \NP, \PSpace) can be reduced to it by a deterministic logarithmic-space (resp. polynomial-time) algorithm. Condition (i) is called {\em membership\/} and condition (ii) {\em hardness}.

\section{Opacity}
  As explained in the introduction, up to one exception, we study in the sequel the complexity of verification of current-state opacity, the definition of which we now recall.

  \begin{defn}[Current-state opacity]
    Let $G=(Q,\Sigma,\delta,I)$ be a DES, $P\colon\Sigma^*\to \Sigma_o^*$ a projection, $Q_S \subseteq Q$ a set of secret states, and $Q_{NS} \subseteq Q$ a set of non-secret states. System $G$ is {\em current-state opaque\/} if for every string $w$ such that $\delta(I,w)\cap Q_S \neq \emptyset$, there exists a string $w'$ such that $P(w)=P(w')$ and $\delta(I,w')\cap Q_{NS} \neq \emptyset$.
  \end{defn}

  The exception is language-based weak opacity. Language-based (weak) opacity is defined over a set of secret behaviors. We recall the most general definition by \cite{Lin2011}.
  
  \begin{defn}[Language-based opacity]
    Let $G=(Q,\Sigma,\delta,I)$ be a DES, $P\colon\Sigma^*\to \Sigma_o^*$ a projection, $L_S \subseteq L(G)$ a secret language, and $L_{NS} \subseteq L(G)$ a non-secret language. System $G$ is {\em language-based opaque\/} if $L_S \subseteq P^{-1}P(L_{NS})$.
  \end{defn}

  Informally, the system is language-based opaque if for any string $w$ in the secret language, there is a string $w'$ in the non-secret language with the same observation $P(w)=P(w')$. In this case, the intruder cannot conclude whether the secret string $w$ or the non-secret string $w'$ has occurred. 
  
  The system is language-based weakly opaque if {\em some\/} strings from the secret language are confused with some strings from the non-secret language. 

  \begin{defn}[Language-based weak opacity]
    Let $G=(Q,\Sigma$, $\delta,I)$ be a DES, $P$ a projection, $L_S \subseteq L(G)$ a secret language, and $L_{NS} \subseteq L(G)$ a non-secret language. System $G$ is {\em language-based weakly opaque\/} if $L_S \cap P^{-1}P(L_{NS}) \neq \emptyset$.
  \end{defn}

  It is worth mentioning that the secret and non-secret languages are often considered to be regular, since for non-regular languages, e.g., for deterministic context-free languages, the inclusion problem is undecidable. \cite{AsveldN00} give a broader picture on (un)decidability of the inclusion problem.

\section{Language-based Weak Opacity}
  \cite{Lin2011} has shown that deciding language-based weak opacity is polynomial for the secret and non-secret languages given by finite automata. The idea is based on the observation that $L_S \cap P^{-1}P(L_{NS}) \neq \emptyset$ if and only if $P(L_{S})\cap P(L_{NS}) \neq \emptyset$, where the later can be checked in polynomial (quadratic) time by representing $P(L_i)$, $i\in\{S,NS\}$, as an NFA (with $\eps$-transitions), computing the product of such NFAs, and deciding non-emptiness of the resulting automaton.
  
  However, is the problem the hardest problem in the class of polynomially solvable problems? Equivalently, is the problem of deciding language-based weak opacity \PTime-complete? If it were, its verification would probably not be parallelizable. Notice the word ``probably'' referring to the longstanding open problem from complexity theory similar to the famous problem whether \PTime = \NP. We now show that the problem is \NL-complete and, consequently, can be efficiently solved on a parallel computer, see \cite{AroraBarak2009}.
  
  \begin{thm}\label{thm_sdnl-c}
    Deciding language-based weak opacity for a DES where both the secret language $L_{S}$ and the non-secret language $L_{NS}$ are modeled by NFAs is \NL-complete.
  \end{thm}
  \begin{pf}
    \cite{Lin2011} has shown that $L_S \cap P^{-1}P(L_{NS}) \neq \emptyset$ is equivalent to $P(L_{S})\cap P(L_{NS}) \neq \emptyset$. Let $L_{S}$ and $L_{NS}$ be represented by NFAs $G_1=(Q_1,\Sigma,\delta_1,Q_{0,1},Q_{m,1})$ and $G_2=(Q_2,\Sigma,\delta_2,Q_{0,2}$, $Q_{m,2})$, respectively. To check that $P(L_{S})\cap P(L_{NS}) \neq \emptyset$ is satisfied, the NL algorithm guesses two pairs of states $(q_{0,1},q_{0,2})\in Q_{0,1}\times Q_{0,2}$ and $(q_{m,1},q_{m,2})\in Q_{m,1}\times Q_{m,2}$ and verifies that the pair $(q_{m,1},q_{m,2})$ is reachable from $(q_{0,1},q_{0,2})$ in the product automaton. For more details how to check reachability in NL, the reader is referred to \cite{Masopust2018}.
    
    To show \NL-hardness, we reduce the {\em DAG reachability problem}: given a directed acyclic graph $G=(V,E)$ and two vertices $s,t\in V$, the problem asks whether vertex $t$ is reachable from vertex $s$. 
    From $G$, we construct an NFA $A=(V\cup\{t'\},\{a,b\},\delta,s,V\cup\{t'\})$, where $a$ is an observable and $b$ an unobservable event, and for every edge $(p,r)\in E$, we add the transition $(p,a,r)$ to $\delta$. Moreover, we add a new state $t'$ and a new transition $(t,b,t')$. Let $L_S$ be the language of the automaton $(V\cup\{t'\},\{a,b\},\delta,s,\{t\})$ and $L_{NS}$ the language of the automaton $(V\cup\{t'\},\{a,b\},\delta,s,\{t'\})$. Obviously, $L_{S}$ is nonempty if and only if $L_{NS}$ is nonempty, which is if and only if $t$ is reachable from $s$. Then, if $a^k$ is the label of transitions from $s$ to $t$, then $P(a^k)=a^k \in P(L_{S})$ and $P(a^kb)=a^k\in P(L_{NS})$. Hence $P(L_{S})\cap P(L_{NS})\neq\emptyset$ if and only if $t$ is reachable from $s$. 
  \hfill$\qed$\end{pf}

  We point out that using a unique observable event for every transition can show \NL-hardness for DESs modeled as DFAs.

\section{Opacity Verification}
  In the section, we discuss several structural restrictions on the systems and the effect of these restrictions on the complexity of verification of current-state opacity. Namely, we discuss the restriction on the number of observable and unobservable events, then we combine this restriction with the requirement on acyclicity of the system model, and finish the section by relaxing the restriction on acyclicity to allow deadlock freeness.

  To simplify the proofs, we first reduce current-state opacity to the language inclusion problem. This reduction is similar to that of \cite{WuLafortune2013} reducing current-state opacity to language-based opacity.
  \begin{lem}\label{CSOinclusion}
    Let $G=(Q,\Sigma,\delta,I)$ be a DES, $P\colon\Sigma^*\to \Sigma_o^*$ a projection, and $Q_S, Q_{NS} \subseteq Q$ sets of secret and non-secret states, respectively. Let $L_S$ denote the marked language of the automaton $G_{S}=(Q,\Sigma,\delta,I,Q_S)$ and $L_{NS}$ denote the marked language of the automaton $G_{NS}=(Q,\Sigma,\delta,I,Q_{NS})$. Then $G$ is current-state opaque if and only if $P(L_{S})\subseteq P(L_{NS})$.\footnote{\baselineskip0\baselineskip Here, $P(L_S)$ is understood as being represented by an NFA with $\eps$-transitions obtained from $G_S$ by replacing transition $(p,a,q)$ with $(p,P(a),q)$; analogously for $P(L_{NS})$.}
  \end{lem}
  \begin{pf}
    Assume that $w$ is such that $\delta(I,w)\cap Q_S \neq \emptyset$. This is if and only if $P(w)\in P(L_{S})$. Then, by definition, there is a string $w'$ such that $P(w)=P(w')$ and $\delta(I,w')\cap Q_{NS} \neq \emptyset$, which is if and only if $P(w)\in P(L_{NS})$.
  \hfill$\qed$\end{pf}

  Furthermore, \cite{Cassez2012} pointed out that the verification of current-state opacity is at least as hard as deciding universality. Indeed, for a DES $G=(Q,\Sigma,\delta,I,F)$, $L(G)=\Sigma^*$ if and only if $G$ is current-state opaque with respect to $Q_S=Q\setminus F$ and $Q_{NS}=F$. 

  This observation and Lemma~\ref{CSOinclusion} together with the results on the complexity of deciding universality and inclusion give us strong tools to show lower and upper complexity bounds for deciding (current-state) opacity.

\subsection{Restriction on the number of events}
  Our first restriction concerns the number of observable and unobservable events in the system. The following result thus improves the general case in two ways: (i) compared to the general settings where more than a single initial state is allowed, although the transitions are all deterministic, we allow only a single initial state, and hence keep the system deterministic, and, mainly, (ii) we restrict the number of observable events to two and the number of unobservable events to one.
  \begin{thm}\label{thm2}
    Deciding current-state opacity for a DES modeled by a DFA with three events, one of which is unobservable, is \PSpace-complete.
  \end{thm}
  \begin{pf}
    Membership in \PSpace was shown by~\cite{Saboori2011}, and also follows directly from Lemma~\ref{CSOinclusion}.

    To show hardness, we reduce the DFA-union universality problem shown to be \PSpace-complete by \cite{Kozen77}. Thus, let $\A_1,\ldots, \A_n$ be DFAs over the alphabet $\Sigma=\{0,1\}$. We let both events of $\Sigma$ be observable. Without loss of generality, we may assume that the initial state of $\A_i$, for $i=1,\ldots,n$, is not reachable from any other state.\footnote{\baselineskip0\baselineskip Otherwise, we can modify $\A_i$ by adding a new state $q'_{0,i}$ that is marked if and only if the initial state $q_{0,i}$ is marked, add, for every transition $(q_{0,i},e,p)$, a new transition $(q_{0,i}',e,p)$, and let $q_{0,i}'$ be the only initial state. This modification does not change the language of $\A_i$. Moreover, we put $q_{0,i}'$ to the set $Q_S$, $Q_{NS}$ or $Q\setminus (Q_S\cup Q_{NS})$ to which $q_{0,i}$ belongs, which preserves the opacity property.}
    Let $G$ denote the nondeterministic union of all $\A_i$'s, that is, $L(G)=\bigcup_{i=1}^{n} L(\A_i)$. \cite{Kozen77} showed that deciding whether $L(G) = \Sigma^*$ is a \PSpace-hard problem, and hence deciding current-state opacity of $G$ is \PSpace-hard by the observation of~\cite{Cassez2012} formulated below Lemma~\ref{CSOinclusion}.
  
    Notice that although the transitions of $G$ are deterministic, $G$ may have more than a single initial state, say $I=\{q_1,\ldots,q_n\}$. We now further modify $G$ by adding a new unobservable event $a$ and the transitions $(q_i,a,q_{i+1})$, for $i=1,\ldots,n-1$, and let $q_1$ be the sole initial state. Denoting the result by $G'$, we can see that $G'$ is a DFA, and that the observers of $G$ and $G'$ coincide; indeed, the initial state of the observer of $G$ is $I$, because both events of $G$ are observable, and the initial state of the observer of $G'$ is the set of states reachable from $q_1$ under the sequences of unobservable event $a$, that is, it is $I$ as well. Notice that here we needed the assumption that the initial state of $\A_i$ is not reachable from other states of $\A_i$; otherwise, the observers of $G$ and $G'$ could be different. Altogether, $G$ is opaque if and only if $G'$ is opaque, which completes the proof.
  \hfill$\qed$\end{pf}

  Notice that an unobservable event in the previous theorem is unavoidable because any DFA with all events observable is always in a unique state, and therefore never opaque. However, the reader may wonder what happens if we further restrict the number of observable events to one. We now show that having only one observable event makes the problem computationally easier unless \coNP = \PSpace. This result holds even without any restriction on the number of unobservable events, and for nondeterministic automata.
  
  \begin{thm}\label{thm3}
    Deciding current-state opacity of a DES modeled by an NFA with a single observable event is \coNP-complete.
  \end{thm}
  \begin{pf}
    Membership in \coNP follows from Lemma~\ref{CSOinclusion} and the fact that inclusion for unary NFAs is \coNP-complete, and hardness follows from the complexity of deciding universality for unary NFAs. For both the claims used here, the reader is referred to \cite{StockmeyerM73}.
  \hfill$\qed$\end{pf}

\subsection{Restriction on the structure -- acyclic automata}
  The previous results show that only restricting the number of events does not lead to tractable complexity. But it gives rise to another question whether there are structurally simpler systems for which the opacity verification problem is tractable. 
  
  Structurally the simplest systems we could think of are acyclic DFAs with full observation, recognizing only finite languages. However, these systems are never opaque, since they are deterministic and fully observed. 
  Nontrivial structures to be considered could thus be acyclic NFAs that still recognize only finite languages, and hence do not possess the ability to express all strings over the alphabet. We combine this restriction with the restriction on the number of events.

\begin{thm}\label{thm7}
  Deciding current-state opacity of a DES modeled by an acyclic NFA with at least two observable events is \coNP-complete.
\end{thm}
\begin{pf}
  Assume that the acyclic NFA has $n$ states. Then any string from its language is of length at most $n-1$. Thus, to show that the system is not opaque, an \NP algorithm guesses a subset of secret states and a string of length at most $n-1$ and verifies, in polynomial time, that the guessed subset is reachable by the guessed string. This shows that verifying opacity is in \coNP. Notice that membership in \coNP can also be directly derived from Lemma~\ref{CSOinclusion} and the complexity of inclusion for so-called rpoNFAs of \cite{mfcs16:mktmmt_full} that are more general than acyclic NFAs.
  
  To show \coNP-hardness, we reduce the complement of CNF satisfiability.\footnote{\baselineskip0\baselineskip A (boolean) formula consists of variables, operators conjunction, disjunction and negation, and parentheses. A formula is satisfiable if there is an assigning of {\it true} and {\it false} to its variables making it {\it true}. A literal is a variable or its negation. A clause is a disjunction of literals. A formula is in conjunctive normal form (cnf) if it is a conjunction of clauses; e.g., $\varphi = (x\lor y \lor z) \land (\neg x\lor y \lor z)$ is a formula in cnf with two clauses $x\lor y \lor z$ and $\neg x\lor y \lor z$. Given a formula in cnf, the CNF satisfiability problem asks whether the formula is satisfiable. The formula $\varphi$ is satisfiable for, e.g., $(x,y,z)=(0,1,0)$.} The proof is based on the construction showing that non-equivalence for regular expressions with operations union and concatenation is \NP-complete even if one of them is of the form $\Sigma^n$ for some fixed $n$, see \cite{Hunt73} or \cite{StockmeyerM73}. 
  
  Let $\{x_1,\ldots,x_n\}$ be a set of variables and $\varphi = \varphi_1 \land \cdots \land \varphi_m$ be a formula in CNF, where every $\varphi_i$ is a disjunction of literals. Without loss of generality, we may assume that no clause $\varphi_i$ contains both $x$ and $\neg x$. Let $\neg \varphi$ be the negation of $\varphi$ obtained by de Morgan's laws. Then $\neg\varphi = \neg\varphi_1 \lor \cdots \lor \neg\varphi_m$ is in disjunctive normal form. 
  
  For every $i=1,\ldots,m$, we define a regular expression $\beta_i = \beta_{i,1}\beta_{i,2}\cdots\beta_{i,n}$, where
  \[
    \beta_{i,j} = \left\{
      \begin{array}{cl}
        (0+1) & \text{ if neither } x_j \text{ nor } \neg x_j \text{ appear in } \neg\varphi_i\\
        0   & \text{ if } \neg x_j \text{ appears in } \neg\varphi_i\\
        1   & \text{ if } x_j \text{ appears in } \neg\varphi_i
      \end{array}
      \right.
  \]
  for $j=1,\ldots,n$. Let $\beta = \bigcup_{i=1}^{m} L(\beta_{i})$ be the union of languages defined by expressions $\beta_i$. Then we have that $w\in L(\beta)$ if and only if $w$ satisfies some $\neg\varphi_i$. That is, we have that $L(\beta) = \{0,1\}^n$ if and only if $\neg\varphi$ is a tautology, which is if and only if $\varphi$ is not satisfiable. Notice that the length of every string recognized by $\beta_{i}$ is exactly $n$.

  Let $M$ be an NFA consisting of $m$ paths of length $n$, each corresponding to the language of $\beta_i$, and make the last state of each of these paths non-secret, that it, it is placed to $Q_{NS}$. In addition, add a path consisting of $n+1$ states $\{\alpha_0,\alpha_1,\ldots,\alpha_{n}\}$ and transitions $(\alpha_\ell,a,\alpha_{\ell+1})$, for $0 \le \ell < n$, where $a\in \{0,1\}$. Let $\alpha_n$ be the sole secret state, i.e., $Q_S=\{\alpha_n\}$. Notice that the language of $M$ marked by the states in $Q_S$ is $\{0,1\}^n$, whereas the language marked by the states in $Q_{NS}$ is $L(\beta)$. By Lemma~\ref{CSOinclusion}, $M$ is current-state opaque if and only if $\{0,1\}^n \subseteq L(\beta)$, which is if and only if $\varphi$ is not satisfiable. This completes the proof of \coNP-completeness.
\hfill$\qed$\end{pf}

  Again, we can show that the situation is computationally simpler if only one observable event is allowed.
\begin{thm}\label{thm7b}
  Deciding current-state opacity of a DES modeled by an acyclic NFA with a single observable event is \NL-complete, and hence solvable in polynomial time.
\end{thm}
\begin{pf}
  Membership in \NL follows from Lemma~\ref{CSOinclusion} and the complexity of inclusion for unary languages, see \cite{mfcs16:mktmmt_full}.

  To prove \NL-hardness, we reduce the DAG-reachability problem. Let $G$ be a directed acyclic graph with $n$ vertices, and let $s$ and $t$ be two vertices of $G$. We define an acyclic NFA $\A$ as follows. With each node of $G$, we associate a state in $\A$. Whenever there is an edge from $i$ to $j$ in $G$, we add a transition $(i,a,j)$ to $\A$. The resulting automaton $\A$ is an acyclic NFA. Let $t$ be the sole secret state, i.e., $Q_{S}=\{t\}$, and let $Q_{NS}$ be empty. Obviously, $\A$ is not current-state opaque if and only if there is a string $w \in \{a\}^*$ such that $\delta(s,w) \cap Q_S \neq \emptyset$. Hence $\A$ is not current-state opaque if and only if $t$ is reachable from $s$ in $G$.
\hfill$\qed$\end{pf}

\begin{rem}
  Notice that the choice of $Q_{NS}=\emptyset$ reduces current-state opacity to non-reachability of states from $Q_{S}$. Since this is independent on the automata models, recent results by \cite{CzerwinskiLLLM19} on the lower-bound complexity of reachability in Petri nets gives that deciding current-state opacity is not elementary for Petri nets.\footnote{For the notion of elementary complexity, we refer to the referenced paper.}
\end{rem}

\subsection{Restriction on the structure -- deadlock-free automata}
  Above, we considered systems generating only finitely many behaviors. However, real-world systems are usually not that simple and often require additional properties, such as deadlock freeness. Therefore, we now consider a kind of automata where all cycles are only in the form of self-loops. Such automata are, in our opinion, structurally the simplest deadlock-free DES. Their mark languages form a subclass of regular languages strictly included in {\em star-free languages}, see \cite{BrzozowskiF80} and \cite{SchwentickTV01}. Star-free languages are languages definable by {\em linear temporal logic\/} that is often used as a specification language in automated verification. 

  We now formalize our model. Let $\A=(Q,\Sigma,\delta,I,F)$ be an NFA. The reachability relation $\le$ on the state set $Q$ is defined by $p\le q$ if there is $w\in \Sigma^*$ such that $q\in \delta(p,w)$. The NFA $\A$ is {\em partially ordered (poNFA)\/}  if the reachability relation $\le$ is a partial order. If $\A$ is a partially ordered DFA, we use the notation {\em poDFA}.

  We then immediately obtain the following result for nondeterministic partially ordered automata.
  \begin{thm}\label{thm89}
    Deciding current-state opacity of a DES modeled by a poNFA with only two events, both of which are observable, is \PSpace-complete.
  \end{thm}
  \begin{pf}
    Membership in \PSpace follows from Lemma~\ref{CSOinclusion} and the results on the complexity of inclusion for poNFAs, and hardness from the fact that deciding universality for poNFAs with only two events is \PSpace-complete. For both claims see \cite{mfcs16:mktmmt_full}.
  \hfill$\qed$\end{pf}

  The situation is again easier if the model has only a single observable event.
  \begin{thm}\label{thm90}
    Deciding current-state opacity of a DES modeled by a poNFA with a single observable event is \NL-complete.
  \end{thm}
  \begin{pf}
    Membership in NL follows from Lemma~\ref{CSOinclusion} and the corresponding complexity of inclusion, and hardness from the fact that deciding universality for unary poNFAs is \NL-complete, see \cite{mfcs16:mktmmt_full}.
  \hfill$\qed$\end{pf}

  We now consider DES modeled by poDFAs. Since every DFA with all events observable is always in a unique state, and hence never opaque, some unobservable events are necessary to ensure opacity. We show that even one unobservable event makes the opacity verification \PSpace-complete. Consequently, the problem is hard for basically all practical cases.
  
  \begin{thm}\label{thm10}
    Deciding current-state opacity for systems modeld by poDFAs over an alphabet with three events, one of which is unobservable, is \PSpace-complete.
  \end{thm}
  \begin{pf}
    Membership in \PSpace follows from Lemma~\ref{CSOinclusion} and the corresponding complexity of inclusion. 
    
    Let $\A=(Q,\{0,1\},\delta,I,F)$ be a poNFA. By Theorem~\ref{thm89}, deciding current-state opacity for poNFAs with two events, both observable, is \PSpace-complete. From $\A$, we now construct a poDFA $\D=(Q\cup Q',\{0,1,a\},\delta',s,F)$ by 'determinizing' it with the help of new events that we then encode in unary. In more detail, for every state $p$ with two transitions $(p,x,r)$ and $(p,x,q)$ with $p\neq q$, we replace the transition $(p,x,q)$ with two transitions $(p,x',p')$ and $(p',x,q)$, where $x'$ is a new event and $p'$ a new state (added to $Q'$); see Fig.~\ref{fig2} for an illustration.
    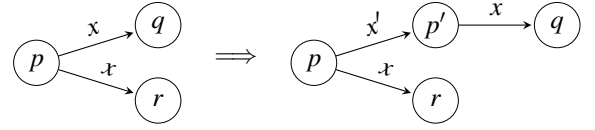
\begin{figure}
      \centering
      \begin{tikzpicture}[baseline,->,>=stealth,auto,shorten >=1pt,node distance=1.6cm,
        state/.style={circle,minimum size=6mm,inner sep=1,very thin,draw=black,initial text=}]
        \node[state]  (1) {$p$};
        \node         (0) [right of=1] {};
        \node[state]  (2) [above of=0,node distance=.5cm] {$q$};
        \node[state]  (3) [below of=0,node distance=.5cm] {$r$};
        \path
          (1) edge node[pos=.7,sloped] {$x$} (2)
          (1) edge node[pos=.4,sloped] {$x$} (3)
          ;
      \end{tikzpicture}
      \quad $\Longrightarrow$ \quad
      \begin{tikzpicture}[baseline,->,>=stealth,auto,shorten >=1pt,node distance=1.6cm,
        state/.style={circle,minimum size=6mm,inner sep=1,very thin,draw=black,initial text=}]
        \node[state]  (1) {$p$};
        \node         (0) [right of=1] {};
        \node[state]  (2) [above of=0,node distance=.5cm] {$p'$};
        \node[state]  (4) [right of=2] {$q$};
        \node[state]  (3) [below of=0,node distance=.5cm] {$r$};
        \path
          (1) edge node[pos=0.8,sloped] {$x'$} (2)
          (2) edge node[pos=0.5,sloped] {$x$} (4)
          (1) edge node[pos=0.4,sloped] {$x$} (3)
          ;
      \end{tikzpicture}
      \caption{The 'determinization'; $x'$ and $p'$ are a new event and a new state}
      \label{fig2}
    \end{figure}
    
    In this way, we eliminate all nondeterministic transitions. The automaton is now deterministic with the set of initial states $I$. Let $\Gamma$ be the set of all the new events we created by the construction. Notice that the number of these events is bounded by the number of edges in the original automaton, and hence polynomial. 
    
    We now show how to encode the new events as strings over a unary alphabet $\{a\}$. Let $m=|\Gamma|$ be a number of new events and $\text{enc}\colon \Gamma \to \{a,aa, \ldots, a^{m}\}$ be an arbitrary encoding (injection). We replace every transition $(p,x',p')$, for $x'\in\Gamma$, by the sequence of transitions $(p,\text{enc}(x'),p')$, which requires to add up to $m$ new states to $Q'$. 
    For instance, if $(p,x,p')$ and $(p,y,p'')$ are two transitions with $x,y\in\Gamma$, and $\text{enc}(x)=aa$ and $\text{enc}(y)=aaa$, then $(p,x,p')$ is replaced by transitions $(p,a,p_1), (p_1,a,p')$, where $p_1$ is a new state added to $Q'$, and $(p,y,p'')$ is replaced by transitions $(p,a,p_1), (p_1,a,p'), (p',a,p'')$; see Fig.~\ref{fig6}. Notice that we have not set the secret status of states of $Q'$, and hence we have that the states of $Q'$ are neither secret nor non-secret.
    \begin{figure}
      \centering
      \begin{tikzpicture}[baseline,->,>=stealth,auto,shorten >=1pt,node distance=1.3cm,
        state/.style={circle,minimum size=6mm,inner sep=1,very thin,draw=black,initial text=}]
        \node[state]  (1) {$p$};
        \node         (0) [right of=1] {};
        \node[state]  (2) [above of=0,node distance=.5cm] {$p'$};
        \node[state]  (3) [below of=0,node distance=.5cm] {$p''$};
        \path
          (1) edge node[pos=.7,sloped] {$x$} (2)
          (1) edge node[pos=.2,sloped] {$y$} (3)
          ;
      \end{tikzpicture}
      \quad $\Longrightarrow$ \quad
      \begin{tikzpicture}[baseline,->,>=stealth,auto,shorten >=1pt,node distance=1.3cm,
        state/.style={circle,minimum size=6mm,inner sep=1,very thin,draw=black,initial text=}]
        \node[state]  (1) {$p$};
        \node[state]  (2) [right of=1] {$p_1$};
        \node[state]  (3) [right of=2] {$p'$};
        \node[state]  (4) [right of=3] {$p''$};
        \node (0) [right of=2] {};
        \path
          (1) edge node {$a$} (2)
          (2) edge node {$a$} (3)
          (3) edge node {$a$} (4)
          ;
      \end{tikzpicture}
      \caption{The encoding $\text{enc}(x)=aa$ and $\text{enc}(y)=aaa$}
      \label{fig6}
    \end{figure}
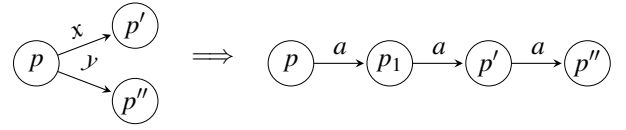
    
    Assume that $I=\{q_1,\ldots,q_{n}\}$. To obtain a single initial state (labeled by $q_0'$), we add, for $i=1,\ldots,n$, a new state $q_i'$ and two transitions $(q_{i-1}',a,q_i')$ and $(q_i',0,q_i)$ as depicted in Fig.~\ref{fig4}. The resulting automaton, $\D$, is a poDFA over the alphabet $\{0,1,a\}$ with polynomially many new events and states, and a single initial state $q_0'$. 
    \begin{figure}
      \centering
      \begin{tikzpicture}[baseline,auto,->,>=stealth,shorten >=1pt,node distance=1.3cm,
        state/.style={ellipse,minimum size=6mm,inner sep=0pt,very thin,draw=black,initial text=},
        every node/.style={font=}]
        \node[state,initial]  (1) {$q_0'$};
        \node[state]  (2) [right of=1]  {$q_1'$};
        \node[state]  (3) [right of=2]  {$q_2'$};
        \node[state]  (4) [right of=3]  {$q_3'$};
        \node[state]  (5) [right of=4]  {$q_4'$};
        \node[state]  (6) [below of=2,node distance=1.8cm]  {$q_1$};
        \node[state]  (7) [below of=3,node distance=1.8cm]  {$q_2$};
        \node[state]  (8) [below of=4,node distance=1.8cm]  {$q_3$};
        \node[state]  (9) [below of=5,node distance=1.8cm]  {$q_4$};
        
        \path
          (1) edge node {$a$} (2)
          (2) edge node {$a$} (3)
          (3) edge node {$a$} (4)
          (4) edge node {$a$} (5)
          (2) edge node {$0$} (6)
          (3) edge node {$0$} (7)
          (4) edge node {$0$} (8)
          (5) edge node {$0$} (9)
          ;
          
        \begin{pgfonlayer}{background}
          \path (6.north -| 9.east) + (0.2,0.3)     node (a) {};
          \path (9.south -| 6.west) + (-0.2,-0.3)   node (b) {};
          \path[rounded corners, draw=black] (a)    rectangle (b);
        \end{pgfonlayer}
      \end{tikzpicture}  
      \caption{Construction of an automaton with a single initial state from an automaton with four initial states}
      \label{fig4}
    \end{figure}
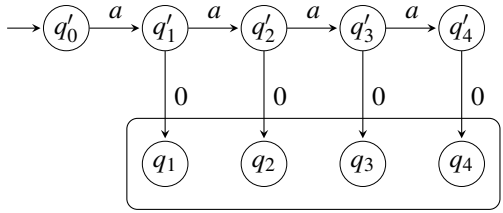

    Let $P$ be the projection from $\{0,1,a\}^*$ to $\{0,1\}^*$, and let $P(\D)$ denote the poNFA obtained from $\D$ by replacing every transition $(p,a,q)$ by $(p,P(a),q)$. Then $\D$ is current-state opaque with respect to $P$ if and only if $P(\D)$ is current-state opaque (with respect to the identity map), which is if and only if $\A$ is current-state opaque (with respect to the identity map).
  \hfill$\qed$\end{pf}

\section{Consequences}\label{secConsequences}
  \cite{WuLafortune2013} provided polynomial reductions among several notions of opacity, including current-state opacity, language-based opacity (for regular languages), initial-state opacity, and initial-and-final-state opacity. Inspecting the reductions, it can be seen that they preserve both acyclicity and partial order. Moreover, eliminating the unnecessary \textit{Trim} operations, we obtain deterministic logarithmic-space reductions, see \cite{Masopust2018} for more details.
  Consequently, the results for current-state opacity also hold for language-based opacity (for regular languages) and initial-and-final-state opacity. Moreover, the lower bounds also hold for $K$-step opacity, since current-state opacity is a special case thereof.

  The only problematic construction of \cite{WuLafortune2013} is the reduction from language-based opacity (LBO) to initial-state opacity (ISO), which requires that the languages are prefix closed. We briefly depict a general reduction from LBO to ISO. 
  
  Let $G=(Q,\Sigma,\delta,I)$ be a DES, $P\colon\Sigma^*\to \Sigma_o^*$ a projection, and $Q_S,Q_{NS} \subseteq I$ sets of secret and non-secret initial states. System $G$ is {\em initial-state opaque\/} if for every $i \in Q_S$ and every $w \in L(G,i)$, there is $j \in Q_{NS}$ and $w' \in L(G,j)$ such that $P(w) = P(w')$.
  
  Let the languages $L_S$ and $L_{NS}$ of LBO be given by nonblocking automata $G_s$ and $G_{ns}$, respectively. Let $x_s$, $x_{ns}$ be two new states, and let $@$ be a new observable event. To every marked state $r$, we add a transition $(r,@,x_s)$ if $r$ is in $G_s$, and $(r,@,x_{ns})$ if $r$ is in $G_{ns}$. Then $L(G_s) =\overline{L_S} \cup L_S@$ and $L(G_{ns})=\overline{L_{NS}} \cup L_{NS}@$. 
  Let $G$ denote $G_s$ and $G_{ns}$ considered as a single automaton, and let $Q_S$ be the initial states of $G_s$, and $Q_{NS}$ the initial states of $G_{ns}$. Then,  $P(L_S)\subseteq P(L_{NS})$ if and only if $P(L(G_s))\subseteq P(L(G_{ns}))$, which is if and only if $G$ is ISO.

  Notice that the reduction does not preserve the number of observable events. This could be solved by encoding the events of $G$ in binary, but then the reduction may not preserve partial order and the number of events if the languages $L_S$ and $L_{NS}$ are unary.
  However, adjusting the results for current-state opacity according to the suggested reduction, we obtain similar results also for initial-state opacity.

\bibliography{mybib}

\end{document}